# The design of high-speed data transmission method for a small nuclear physics DAQ system

Zhou Wenxiong[1,2], Wang Yanyu[1], Nan Gangyang[1,2], Zhang Jianchuan[1,2]

(1. Institute of Modern Physics，Chinese Academy of Sciences，LanZhou，730000，China，

2. University of Chinese Academy of Sciences ，Beijing，100039，China)

**Abstract**: A large number of data need to be transmitted in high-speed between Field Programmable Gate Array (FPGA) and Advanced RISC Machines 11 micro-controller (ARM11) when we design a small data acquisition (DAQ) system for nuclear experiments. However, it is a complex problem to beat the target. In this paper, we will introduce a method which can realize the high-speed data transmission. By this way, FPGA is designed to acquire massive data from Front-end electronics (FEE) and send it to ARM11, which will transmit the data to other computer through the TCP/IP protocol. This paper mainly introduces the interface design of the high-speed transmission between FPGA and ARM11, the transmission logic of FPGA and the driver program of ARM11. The research shows that the maximal transmission speed between FPGA and ARM11 by this way can reach 50MB/s theoretically, while in nuclear physics experiment, the system can acquire data with the speed of 2.2MB/s.

**Keyword**: FPGA; ARM11; high-speed transmission; driver; DAQ;

IPAC: 07.05.Bx,07.05.Hd,28.41.Rc

## 1. Introduction

In order to reflect the mechanism of the nuclear reaction, we have to measure many groups of parameters currently when performing the nuclear physics experiments [1]. Even though it is a small nuclear physics experiment, we still need to acquire several parameters because of the relevance of the particles produced by particles collision in accelerator. So as to meet the needs of the small unclear physics experiment, a DAQ system based on FPGA and ARM11 is developed. In this system, FPGA is applied to control Front-end electronics (FEE), and transmit the data converted by ADC or TDC to ARM11. And then, an ARM11 is designed to send the data to the other computer for real-time storing and processing. Considering the maximal DAQ rate of FEE is 8MB/s, the speed of data transmission from FPGA to ARM11 must be more than 8MB/s. In comparison with several other transmission methods between FPGA and ARM11, we select the high-speed parallel transmission method because of its higher data rate. There are two points in this transmission method: firstly, FPGA is designed as a memory connected on ARM11's IO bus; secondly, FPGA is controlled by SROM controller (SROMC) which is integrated in ARM11. This data transmission mechanism can raise DAQ rate between FPGA and ARM11 up to 50MB/s theoretically, which means that it can meet the design requirement of the whole DAQ system. In the unclear experiment, the system was used to obtain energy spectrum of $Na^{22}$. And the spectrum is the same with that obtained through PHILLIPS7164, which is a CAMAC module.

## 2. The transmission methods between FPGA and ARM11

The communication methods between FPGA and ARM11 include serial communication, custom parallel communication and the high-speed parallel communication [2]. The speed of those communication methods is show in table 1.

Table1: the speed of different communication methods

| Type | Protocol | Speed |
|---|---|---|
| Serial communication | I2C | 156KB/s |
|  | SPI | 1.4MB/s |
| Custom parallel communication | Custom | 0.4MB/s |
| High-speed parallel communication | SROM | 50MB/s |

Owing to limitations of the protocol and characteristics for the device, the serial communication method couldn't meet the least requirement of the DAQ system. For custom parallel communication method, the data and address bus must consist of GPIO ports. And the experiment, which is designed to get the maximal electrical level change rate of GPIO ports in Linux, shows that the minimal period of the electrical level change is 5μs. So if the custom data bus width is 16 bits, the transmission speed can be calculated through the following formula:

$$S = \frac{1}{T} \times W$$

As "W= 16 bits", and T=5μs, the transmission speed(S) is about 0.4MB/s. It still can't meet the requirement of the whole DAQ system.

In comparison, micro-controllers have a special memory bus with a very high speed data exchange rate as its clock frequency is 133MHz and the data bus width is 16 bits. Because each memory access needs a few cycles, the transmission speed can reach about 50MB/s. Overall, the speed of high-speed transmission method is fast, and it can fit the bill of the system.

## 3. The implementation of high-speed parallel communication method

### 3.1. The logic design

The type of ARM11 used in this DAQ system is S3C6410 and the SROM controller (SROMC) integrated in ARM11 can support 6 banks. However, because the demand of the whole system, FPGA controlled by the SROMC of the ARM11 has to be designed as a memory in bank 4[3]. The following figure (Fig1) shows the hardware structure of the DAQ system.

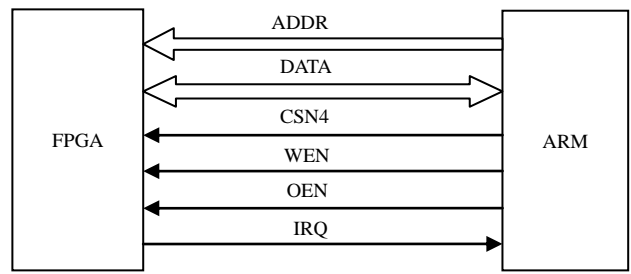

Fig1. Hardware structure of the system

In this structure, a 32KB FIFO whose reading and writing logic is controlled by the SROMC is designed inside FPGA. The connection between ARM11 and FPGA consist of a 16-bit data bus, a 16-bit address bus, reading control signal lines, writing control signal lines and interrupt signal line. Considering the difference of the control logic between the FIFO and the SROMC, the control signal of the SROMC has to be processed in FPGA for data R&W in the FIFO correctly. Following pictures show us the different control logic of the FIFO and the SROMC. Fig2 is a timing block diagram of the FIFO, while Fig3 is the timing block diagram of the SROMC.

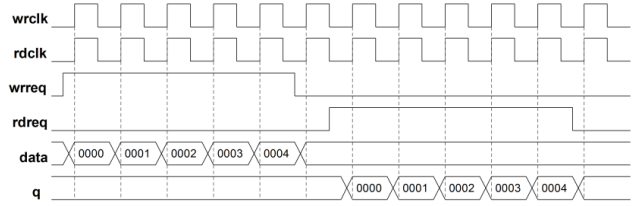

Fig2. The timing block diagram of the FIFO

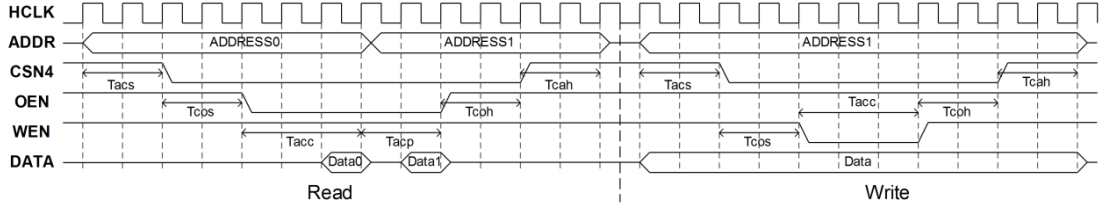

Fig3. The timing block diagram of the SROMC

As we can see from Fig2 and Fig3, when referring to the data input and output, they work in different ways. While the FIFO is controlled by reading and writing clock, the SROMC is controlling through reading and writing control signal. In regard to this, we need to convert the reading and writing control signal of the SROMC into reading and writing clock for the FIFO. After the study and some analysis, the following logic formula is available for conversion.

$$rdclk = \overline{CSN4} \ \& \ \overline{OEN}$$
$$wrclk = \overline{CSN4} \ \& \ \overline{WEN}$$

There is one point that we should pay special attention to: Because of race and hazard in the sequential logic circuit, the rdclk and wrclk signal must be synchronized by the system clock when the logic circuit was adopted to create reading and writing clock for the FIFO. As the system clock frequency of FPGA is 50MHz, the cycle of the rdclk and wrclk signal has to be more than 40ns. For example, supposing the cycle is 40ns, as the frequency of the clock HCLK is 133MHz, we set the value of $T_{acs}$, $T_{cos}$, $T_{coh}$, $T_{cah}$ and $T_{acp}$ to 0 and the value of $T_{acc}$ to 3. Additionally, the SROMC has different control mode, and in its page control mode, $T_{acp}$ is used to read data circularly. So if the SROMC is set to normal mode, $T_{acp}$ will be inactive or ineffective. In such a configuration, every time the SROMC read a 16-bit, we convert the reading and writing signal to reading and writing clock for the FIFO. However, for the reason that the FIFO shares its data bus and address bus with other devices, we should check the CSN4 signal before any reading and writing operation. Otherwise the reading or writing logic will deliver a wrong result and likely the whole system will crash or deliver wrong measurement results. Fig4 is a rdclk signal obtained from FPGA with the help of this logic. The maximal communication rate can reach the value of 50MB/s theoretically, which can commendably meet the need of the DAQ system.

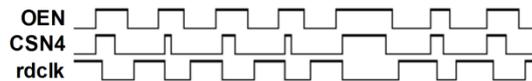

Fig4. Reading clock of FIFO

### 3.2. The driver design for ARM11

It is common that an event generated randomly and casually when we carry on an unclear physics experiment. So the DAQ system couldn't figure out when is the best time to read data. In order to solve this problem, interrupt mechanism is applied in the system to achieve efficient outcomes [4][5]. It works like this: FPGA is set to notify ARM11 to read data when the cumulative number of the event (Multi-Event mode) reaches preset value. In view of the fact that this method can reduce the burden of ARM11, the driver is expected to process the interrupt signal and notify the applications to deal with it. Fig5 shows the driver and the application structure of ARM11.

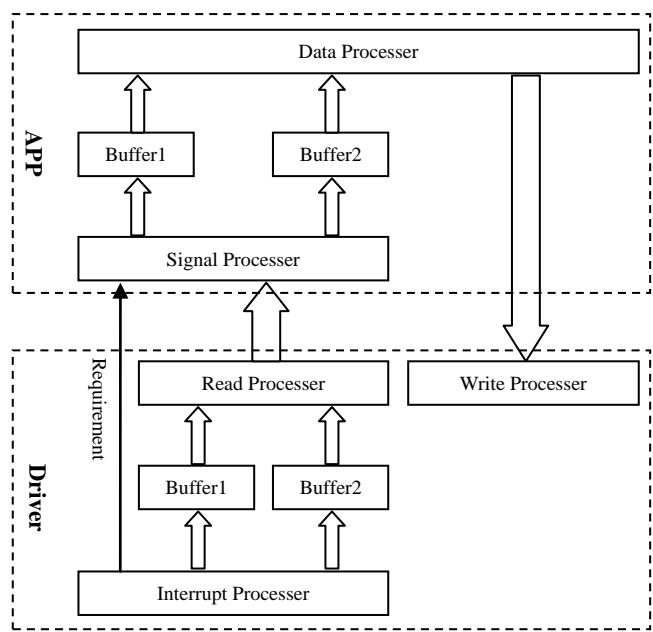

Fig5. Driver and Application structure

The interrupt processer module is used to respond to hardware interrupts [6], and when the number reaches preset value, FPGA will transmit an interrupt message to ARM11, whose interrupt processer module will read the data from FPGA and write it into a buffer. Furthermore, in order to decrease the rate of event losses and the dead time (the time within which a second pulse is not detected), we choose the ping-pong buffer in the driver, which means that even when the application is reading data from one of the two buffers, the interrupt processer module can respond to hardware interrupt, and write data into the other buffer. Besides that, the application of ARM11 is a design based on multi-threading, so the signal and data can be processed in different threads. Specifically speaking, the signal processing thread is occupied to respond to the data reading requirement coming from the driver, and it also used to read data from the driver buffer and write it into the application buffer, while the data processing thread is occupied to analyze the data read by signal processing thread. Similarly, the ping-pong buffer is also applied in application, so that the signal processing module and the data processing module can not only work at the same time, but also reduce the risk of dead time and increase the throughput of the data.

## 4. The experiments and results

The DAQ system achieved by high-speed data transmission method has been tested in following two environments.

**First**: P1010, a NIM module using to generate pulse and gate, was used as the signal source in laboratory to produce two kinds of signals, the pulse signal and the gate signal. The pulse signal generated was sent to the small DAQ system, while the gate signal should be processed by GG8000 which is also a NIM module before it was sent to the system. The experimental results show that the maximal data transmission rate is about 2MB/s, and the frequency of event generation is 250kHz.

**Second:** we chose $Na^{22}$ as the radioactive source, and carry on the experiment in the environments. We converted the energy-signal obtained by $LaBr_3$ crystal detector to the pulse signal and the gate signal after it is amplified by photomultiplier. Then both kinds of signals are sent to the DAQ system. Finally we obtained the following energy spectrum in Fig6. To our relief, the spectrum is the same with what is obtained through PHILLIPS7164, which is a CAMAC module.

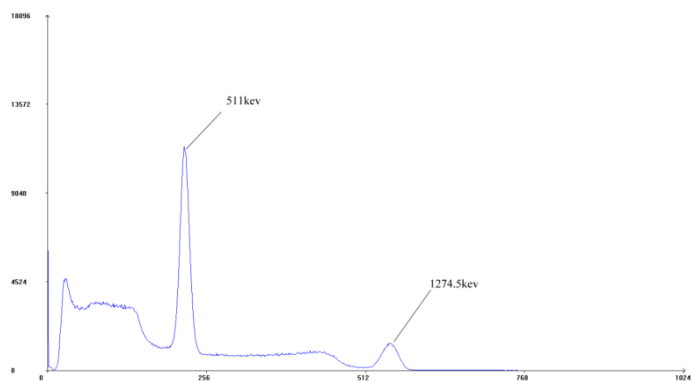

Fig6. The energy spectrum of $Na^{22}$

## 5. conclusion

The high-speed transmission method successfully solved the problem that massive data in a small nuclear DAQ system is difficult to be transmitted in high-speed. The first realistic experimental results show that the maximal data transmission rate can reach the point of 2MB/s, while the rate of event generation is 250kHz. The second energy spectrum test illustrated that this method can meet the need of the small DAQ system. However, as a result of some tiny flaw from the other parts in the system, the whole system can't run at the speed of 8MB/s. once all the parts of the system are completed, the system can run at full speed.

**References:**
[1] Y.Y. Wang, et al. Nuclear Electronics & Detection Technology ,2007, 27(1):18
[2] Y.L. Liao, et al. ARM and FPGA design and application, China electric power press
[3] X.P. Zhu, et al. Computer Engineering and Design, 2009, 30(13):3088
[4] J.C. Zhang, et al. High power laser and particle beams, 2012, 24(11):2727
[5] G.Y. Nan, et al. Nuclear Electronics & Detection Technology, 2011, 31(11):1250
[6] J. Corbet, et al. Linux device drivers, China electric power press